\documentclass[prl]{revtex4}% 
\pdfoutput=1

\usepackage{epsfig}
\usepackage{graphics}

\begin{document}
\bigskip
\hskip 4in\vbox{\baselineskip12pt \hbox{FERMILAB-PUB-09-361-A}  }
\bigskip\bigskip\bigskip

\title{Holographic Noise in Interferometers}

\author{Craig J. Hogan}
\affiliation{  University of Chicago and Fermilab}

%\section{Introduction}
\begin{abstract} %%% Abstract to run on from here.
Arguments based on general principles of quantum mechanics suggest that a minimum length or time associated with   Planck-scale unification may entail a new kind of observable uncertainty in the transverse position of macroscopically separated bodies.  Transverse positions vary randomly about classical geodesics in space and time by about the geometric mean of the Planck scale and separation, on a timescale corresponding to their separation.    An effective theory based on a Planck information flux limit, and normalized by the black hole entropy formula, is developed to  predict measurable correlations, such as the statistical properties of noise in  interferometer signals.  A connection with holographic unification is illustrated by representing Matrix theory position operators with a  Schr\"odinger wave equation, interpreted as a paraxial wave equation with a Planck frequency carrier.  Solutions of this equation are used to derive formulas for the spectrum of beamsplitter position fluctuations and equivalent strain noise in a Michelson interferometer, determined by the Planck time, with no other parameters. The spectral amplitude of equivalent strain derived here is a factor of $\sqrt{\pi}$ smaller than previously published estimates. Signals in two nearly-collocated interferometers are predicted to be highly correlated, a feature that may  provide convincing evidence for or against this interpretation of  holography.

\end{abstract}
\maketitle

\section{Introduction}

Spacetime was invented using  classical concepts, such as positions of bodies and ticks of clocks.  Although classical spacetime serves remarkably well as a backdrop for precise description of microscopic quantum systems, such as atoms and even particle collisions,  
the concept of a spacetime event is not easily interpreted in the
context of quantum mechanics.  For one thing, the notion of a
pointlike event does not take account of the particle/wave duality of
quantum mechanics.  For another, events themselves are not, even in
principle, observable quantities, but are defined only by
interactions of mass-energy.  

A quantum model for measuring the invariant interval between spacelike events using light quanta and clocks can be constructed in a spacetime with one spacelike dimension\cite{wigner,salecker}.  However, as described below, if there is a minimum fundamental wavelength, event positions defined by wave phases  in  two spacelike dimensions--- for example, reflections off a beamsplitter surface in a Michelson interferometer--- are intrinsically uncertain, by  the geometric mean of the minimum wavelength and the event separation.   This paper describes how a new fundamental nonlocal quantum uncertainty of spacetime  position associated with a minimum wavelength at the Planck scale could be studied with  experiments that make precise position measurements in two directions across  macroscopic distances.

It is well known that spacetime must adopt some new quantum character below the Planck length,
$\lambda_P= \sqrt{\hbar G_N/c^3}=1.616 \times 10^{-35}$ meters, or Planck time, $t_P \equiv
\sqrt{\hbar G_N/c^5}= 5.390\times 10^{-44}$ seconds.  No spatially localized quantum particle states can exist in
classical spacetime above the Planck energy, because of
gravity. A  particle spatially localized within a Planck
volume lies within the Schwarzschild radius for its energy, behind an
event horizon where it can never be seen. The interpretation of spatial wavefunctions must be modified in some way at the Planck scale, an effect that may appear as a minimum fundamental wavelength or maximum frequency\cite{kempf}.

The Planck frontier in field theory corresponds to a UV cutoff  in 3+1D spacetime. However, a consistent  theory of mass-energy and spacetime cannot be achieved simply by imposing a   filter on field modes at the Planck scale in some  frame.  A Planck length ruler in some frame will, in some other frame boosted along its length, appear shorter, violating the minimum-length restriction; or well-separated particles in one frame will in some other frame be closer together and form a black hole.  The reconciliation must be achieved by a new unified theory that includes  new, unfamiliar  transformation properties. Moreover, it is recognized that a frequency cutoff entails a violation of Lorentz covariance\cite{mattingly}.  The new effect discussed here violates Lorentz invariance, but in a new way that would not have appeared in current experimental tests.

Remarkably precise insights about unification at  the
Planck scale come from the theory  of  black hole
evaporation,  whereby  a  black hole converts to
quantum particles in flat spacetime.   In particular the   entropy of a black hole, identified with the logarithm of the total number of degrees of freedom of radiated particles, is one quarter of the area of its event horizon in Planck units.  This idea has led to the conjecture\cite{'tHooft:1993gx,Susskind:1994vu,Bousso:2002ju}   that all of physics may be ``holographic'', encoded in some way on two dimensional null surfaces or light sheets with  the  same information content per area as the null surface representing a black hole event horizon.    However, there has been no experimental test of this conjecture.
Although there are candidates (e.g., \cite{Banks:1996vh}) for holographic theories of everything that  incorporate a minimum length at the Planck scale, it is not generally agreed how to apply them to macroscopic experiments operating in  nearly-flat, nearly-classical spacetime.  

In this paper we investigate the experimental consequences of a particular hypothesis about how spacetime emerges in a holographic world.  Relationships between observables at different locations obey a ``Planck information flux'' limit: 
correlations between observables at two events are limited by the capacity of a Planck wavelength null carrier wave between them. Transverse positions then
have an uncertainty described by a simple effective theory based on wave
optics~\citep{Hogan:2007pk,Hogan:2008zw,Hogan:2008ir}.  A
new uncertainty arises because the wavefunction encoding the
transverse component of position spreads by diffraction.  Just as in
diffraction of classical waves, the resulting transverse position
uncertainty of a system has a standard deviation of probability,
\begin{equation}
 \sigma(L) \ge \sqrt{\lambda_0 L/2\pi},
\end{equation}
after longitudinal propagation over a length $L$, if this transverse information is encoded with minimum wavelength $\lambda_0$.  A normalization of the minimum wavelength $\lambda_0 = 2\lambda_P$  is derived from the black hole entropy-area relationship $S = A/4G_N = A/(2 \lambda_P)^2$.

  In this interpretation, families of paths that connect
events and relationships of bodies in classical spacetime have roughly the same status as  rays in optics: they are
an approximate description of a configuration of a physical system of waves.
Rays have a fundamental indeterminacy imposed by diffraction limits;
the actual
physical system, consisting of wave energy (or in this application,
position probability), is  not sharply confined into rays.  A time-averaged classical metric likewise does not capture all of the physical
wavelike qualities.  Just as in wave optics, diffractive blurring and fluctuations of transverse
correlations in such a system can occur on observable scales much larger
than the wavelength (in this application, the Planck length), given a macroscopic propagation distance.

The potentially observable new phenomenon associated with these ideas is a universal  ``holographic noise'': random fluctuations in the relative transverse spacetime positions of bodies widely separated in spacetime, as measured by the phase of null fields\cite{Hogan:2007pk,Hogan:2008zw,Hogan:2008ir}.  The spectral density of this noise is given just by the Planck time, with no parameters aside from numerical factors that depend on a particular experimental setup. The paraxial-wave effective theory  presented here  expresses these ideas quantitatively, and makes  statistical predictions for the effective positions of optical elements and  for the time series of  signals in Michelson interferometers.  The predicted amplitude in equivalent strain units is smaller than the previous published estimate by a factor of $\sqrt{\pi}$. The remaining uncertainties in precise normalization are mainly connected with interpreting the connection of the fundamental wavelength of this effective theory with black hole entropy.

In principle, holographic  noise has a precisely characterized frequency spectrum and an absolutely calibrated normalization from black hole physics, with no parameters.
It also exhibits a new, distinctive  spatial shear  character, unlike metric perturbations that correspond to strain motions, like gravitational waves.      Finally,  signals  in two nearby  interferometers with no physical connection are shown to be  highly correlated. These distinctive signatures can be used to design an apparatus to provide convincing evidence for or against the effect, and to distinguish it from other physical sources of noise. Estimates are given here for the spectral amplitude of correlated displacement as a function of separation of two interferometers.

 This new effect, although quantum-mechanical, does not correspond to quantum fluctuations of gravitons or gravitational radiation,     metric fluctuations derived from noncommutative geometry, or indeed any perturbation of the metric.  It is therefore distinct from previously conjectured  fluctuations of a quantum-gravitational origin (e.g., \cite{Ellis:1983jz,AmelinoCamelia:1999gg,AmelinoCamelia:2001dy,Schiller:2004tf,Ng:2004xr,Smolin:2006pa}),  based on various hypotheses about quantum perturbations of a metric. Holographic noise on macroscopic scales can be described as a new kind of coherent movement of mass-energy relative to an unperturbed 
classical metric--- a fluctuating departure from classical geodesic trajectories.         The position of mass-energy, measured in the transverse direction relative to the separation between  events,   jitters randomly with time relative to  nominal classical geodesics, in an observer-dependent way, on the Planck diffraction scale.  On the other hand, the equivalence principle still holds in a long time average sense, and with respect to different forms of energy.   

Other current tests of Lorentz invariance are not sensitive to this effect. For example, since the theory is built on null sheets, identical  longitudinal   propagation is predicted for  all null particles and radiation, so no dispersion is predicted in arrival times of photons of different energies arriving from cosmological distances, consistent with current limits\cite{fermi2009}. The effect is suppressed for small length scales or  long averaging times. On a subatomic scale of length and time $\lambda_a,t_a$, the holographic spreading over an averaging time $\tau$ is of order $\sqrt{(\lambda_P\lambda_a)(t_a/\tau)}$, below the range of current laboratory tests.

\section{General Arguments for Holographic Uncertainty}

\subsection{Nonlocal comparison of frequency-bounded wavefunctions}

Holographic uncertainty is posited to arise at a primitive level in measurement of spacetime event positions.  It is a general property of measurements of event positions defined by phases of wavefunctions that have a maximum  frequency.  The following general argument shows how it arises  for events corresponding to interactions of   null fields with massive bodies, such as laser light interacting with mirrors of an interferometer. 

Consider a wavefunction $\psi(x,t)$ that represents the quantum-mechanical amplitude for an event to occur at a position $(x,t)$ in  a particular frame.  The wavefunction can represent the state of a photon, say, where $x,t$ represents the location in one spatial dimension.  The position can refer  to distance from a reflecting surface along the null propagation direction. The state has a standard deviation  $\Delta x$ in the spatial domain at a given time, that is, on a spacelike hypersurface in a particular frame.  The same state is described by a wavepacket in the frequency domain with width $\Delta f= \Delta k /2 \pi= c/2\pi \Delta x$.  A frequency cutoff $f_0$ is imposed, so the wavefunction has the character of an envelope of width $\Delta x$  but no fine scale oscillations smaller than $\lambda_0= c/f_0$.  

Suppose this state is used to compare the locations of two interaction events.    The two events are separated by a macroscopic distance $L=ct$.  Relative positions are encoded in the relative phases of the wavefunctions at two events. In a Michelson interferometer, phases of remote reflection events are compared in two different directions.

The frequency spread means that there is a phase uncertainty that accumulates with time or propagation distance.  The phase uncertainty is $\Delta \phi= (2\pi L/\lambda_0) (\Delta f/f_0)$, corresponding to a position uncertainty $\Delta x_\phi= \Delta \phi (\lambda_0/2\pi)= L (\Delta f/f_0)$. Substitution of the wavepacket width $\Delta f$ implies a minimum uncertainty in the wavefunction that describes the quantum mechanical amplitude (or probability distribution) for phase comparison, 
\begin{equation} 
\Delta f_L^2= c f_0/2\pi L, \qquad {\rm or} \qquad \Delta x_L^2= \lambda_0 L/2\pi.
\end{equation}
Note that the computation of this uncertainty does not depend  on  interactions of individual quanta.   It is a general property of waves with a frequency bound, and is not related to quantum fluctuations.  Thus, the calculation can be applied to reflections in an interferometer that involve vast numbers of photons in a coherent state, interacting with a macroscopic patch of a mirror.  On the other hand, the interpretation is the standard quantum-mechanical interpretation of wave mechanics\cite{zurek};  the actual measurement of an event position has some definite and well-defined value even though the system as a whole is described by a wavefunction. Repeated measurements thus show quantum-mechanical noise.  Planck's constant does not appear, because the conjugate variables are both positions in the spatial domain.

This uncertainty was (justifiably) neglected in the analysis of Wigner and Salecker\cite{salecker}, who studied the limits of  measurement of spacetime intervals in models where photons interact with physically realizable quantum-mechanical clocks.  Holographic uncertainty does not involve assumptions about the definition of clocks, but derives from a fundamental limit on the comparison of phases at widely separated points that do not depend on details such as the mass of clocks or other properties of the massive bodies in the interaction.   It is only important if wavefunctions have a maximum frequency, and if event positions at macroscopic separation are  encoded in the relative phase of wavefunctions at two events.  In holographic theories, relative transverse positions may be encoded in Planck-scale phases of states, and in which case they are subject to an unavoidable uncertainty of this magnitude. 

This simple argument suffices to estimate the magnitude of the effect. More detailed arguments presented below suggest particular properties of the uncertainty: the effective motions in holographic theories are only transverse in character and also have a macroscopic transverse coherence. The following examples show how the transverse uncertainty is connected with transverse position measurement and with black hole evaporation states.

 \subsection{Quantum noise in a Planck wavelength interferometer}

Classical spacetime is defined by a metric that describes the intervals between events.  However, neither the spacetime nor the events are directly observable. Measurements require interactions of wavelike quantum particles.

 The limits of the classical picture with a minimum wavelength are illustrated by considering  quantum noise in a Michelson interferometer using single Planck wavelength quanta. We conjecture that there is no operational way to define spacetime event locations in two spacelike dimensions better than this limit, so the same transverse position uncertainty should appear in any apparatus. 

\begin{figure}
\epsfysize=3.2in 
\epsfbox{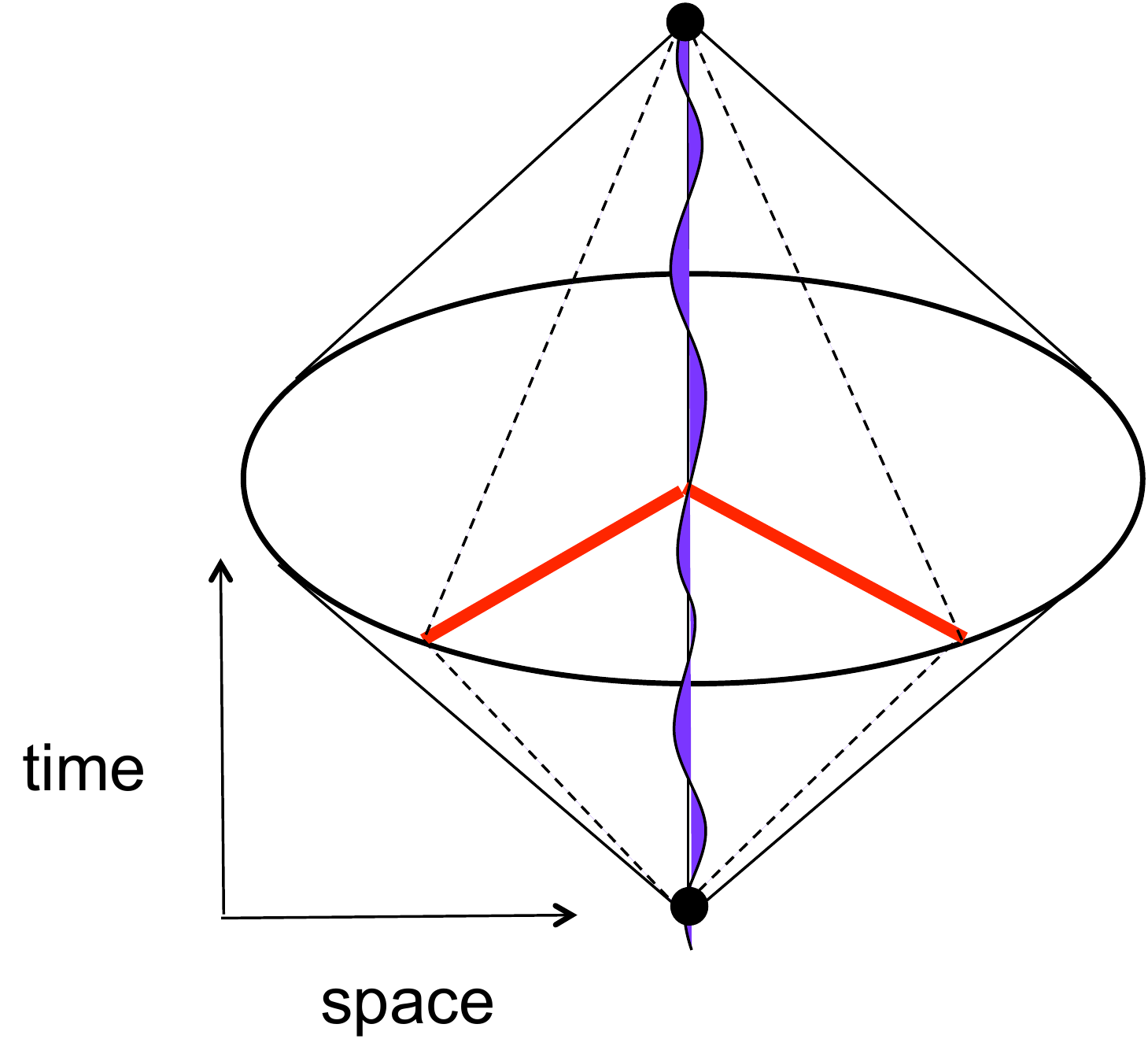} 
\caption{ \label{lightcone}
Spacetime of a Michelson interferometer. The vertical axis corresponds to the classical beamsplitter world line, shown here with schematic noise. The interferometer arms are shown at one time on a spacelike surface where the past and future light cones of two reflection events intersect. Dashed lines show paths of a single particle. The two interaction events of the particle with the beamsplitter (large dots) occur at times separated by $2L/c$.}
\end{figure}

Consider first just the self-interference of a single quantum of wavelength $\lambda_0=h/p_0$. The wave state splits at a beamsplitter (half reflects, half transmits), then propagates along two orthogonal arms of length $L$ and back.  The previously-transmitted part of the state then reflects, and the squared modulus of the summed amplitude is measured at the antisymmetric port.  The two reflections of the  particle state off the beamsplitter  occur at times separated by $2L/c$ (see Figure \ref{lightcone}).  

Two kinds of uncertainty enter into the measured phase signal. The first comes from the position  wavefunction of the reflecting surface of the beamsplitter, with standard deviation $\Delta x$.  The second comes from   the uncertainty $\Delta p$ in the momentum of the reflected particle state. Their standard deviations are related by standard Heisenberg uncertainty\cite{braginsky},
\begin{equation}
\Delta x \Delta p > \hbar/2.
\end{equation}
The first part of the uncertainty in the phase signal comes from the location of the beamsplitter, which affects the path length difference between the two arms. The second  part of the uncertainty in the phase signal is associated with the fact that the different momenta of the waves in the two arms leads to a difference in relative phase when they are recombined and detected. 
When the two parts are recombined their interference   produces a holographic variance in measured phase $\Delta\phi_H^2$ given by the sum in quadrature of the two effects:
\begin{equation}\label{phasevariance}
2 (\lambda_{opt}/2\pi)^2 \Delta\phi_H^2= \Delta x^2 + (2L)^2 (\Delta p/p_0)^2.
\end{equation}
Here we choose to represent the phase in units of a reference wavelength $\lambda_{opt}$, because it will be used below to calculate universal phase noise at optical wavelengths.  The first factor of 2 is from the fact that $\Delta x$ and $\Delta p$   both refer to vectors at 45 degrees to the wavefronts whose phase is measured.

The variance in the measured phase (Eq. \ref{phasevariance})  includes two terms that depend oppositely on $\Delta x$.  It is minimized when the two terms on the right side are equal.  This leads to an uncertainty  in phase,
\begin{equation}\label{Deltaphase}
\Delta\phi_H^2> 2 \pi L \lambda_0/ \lambda_{opt}^2,
\end{equation}
or an
apparent variance $\sigma_X^2$ in the arm length difference $X$,
\begin{equation}\label{Deltax}
\sigma_X^2> L \lambda_0/2\pi.
\end{equation}
If there is a  minimum wavelength
$\lambda_0$, Eq. (\ref{Deltax}) describes an irreducible uncertainty in the phase comparison. In  a time series of measurements this uncertainty leads to the effect we call holographic noise.
We argue below, based on black hole entropy, that the carrier wavelength for fundamental uncertainty should be set to twice the Planck value, $\lambda_0=  2\lambda_P$. 

This system illustrates  several noteworthy features: \begin{enumerate}
\item 
The combined uncertainty is a result of the wavelike character of the measurement. It  is not   there if we time the flights of a classical pointlike particle traveling along rays in a classical background.  The group velocity of a relativistic wavepacket does not change speed with a change in momentum, so a timing measurement of a particle position eigenstate along one axis relative to a perfect clock would not reveal the uncertainty. 
On the other hand, if we use wave cycles to compare the arms,  there is uncertainty.  
\item The transverse character of the position measurement matters.  If the two split parts of the particle state are aligned along the same axis, they share the same momentum perturbation, reducing the phase uncertainty.  Thus, the configuration of the interferometer matters.
\item
If the two arms are not equal, it is the shorter arm that determines the uncertainty, because the accumulated phase uncertainty is less.
\item
Nowhere in the calculation does the mass of the beamsplitter enter explicitly. On the other hand, to achieve the minimum noise, the wavefunction of the beamsplitter surface does matter.  The beamsplitter position wavefunction and the particle momentum uncertainty both contribute to the phase uncertainty.
\item
Nowhere in this calculation does it matter ``where the photon hits the beamsplitter'', that is, the transverse location of the photon beam does not matter to first order.  There is no reference to a ray or a propagation direction, only to the locations of wavefronts and optical surfaces. The noise is thus not due to a jitter in pointing, or the uncertainty in where the particle reflects off the beamsplitter; in fact, this is not measured. It is the position of the wavefronts, and the position of the beamsplitter itself at the times of the two reflections, that affect the measured phase.  
 \end{enumerate}

With coherent radiation including many quanta, the standard quantum uncertainty can be reduced. With $N$ quanta in the same state, the position uncertainty becomes $\Delta x> h/2N \Delta p$, because all the quanta participate in a single measurement with the same $p$. The phase term remains $2L \Delta p /p$ as before because the phases of the coherent quanta are the same. The result is a smaller minimum uncertainty in the apparent arm length difference,
\begin{equation}\label{DeltaxN}
\sigma_X^2>L \lambda_0 /2\pi N.
\end{equation}
A laboratory interferometer uses this feature to reduce the quantum uncertainty in measurement, using enormous numbers of photons in an optical cavity.\cite{caves81} 

However,  Planck radiation cannot have an occupation number greater than unity. (If it did, then the particles would form a black hole in some frame.) In this case the phase of each quantum is independent. For a measurement with $N$ independent quanta,
the first uncertainty becomes $\Delta x >h/N^{1/2} \Delta p$, because   each  quantum makes an independent measurement and the accumulated perturbations add in quadrature. The second becomes  $ 2L N^{1/2} \Delta p /p$ because  the mean phase displacements add in quadrature.  The minimum uncertainty is not reduced in this case, so the Planck interferometer has an irreducible uncertainty in measured phase, corresponding to a variance in  beamsplitter position wavefunction (or arm-length difference) given by Eq. (\ref{Deltax}).

We have discussed this effect as if it is a standard Heisenberg position uncertainty created by Planck waves in  a classical background, with an extra   limit imposed on phase density of quanta. It describes the best measurement possible of relative positions in two spacelike dimensions, given the limitations of interactions of mass-energy with a Planck frequency cutoff.
Wigner and Salecker\cite{wigner,salecker} showed that a spacetime consisting of one time dimension and one spacelike dimension can be consistently defined with quantum-mechanical measurements. The interval between any pair of spacelike-separated events can be measured using clocks on the worldlines of the two events, and light quanta travelling from one worldline to the other.  However, the calculation we have just done suggests that there is more to the story if we add a second spacelike dimension.  If the measurement is done with wavelike quanta, there is additional uncertainty, at the diffraction scale for the waves, in event positions transverse to the light whose wave phase is used for the measurement. The world-line of a beamsplitter defined this way has additional quantum noise.

If an upper  limit on   frequencies and spatial wavenumbers is connected with the unification of spacetime and mass-energy,
it could lead to a fundamental limit on the definition of positions  of mass-energy within a  spacetime. A measurable phase uncertainty could result from a new kind of uncertainty principle due to unification physics. 
 This argument suggests that interferometers may offer a route to Planck scale physics because the uncertainty is magnified in a macroscopic apparatus to much larger than the Planck length.
 Optical-wavelength interferometers can reach the Planck quantum limit because they can measure the mean position of   massive bodies to this precision over a large  surface area, macroscopic on a scale of millimeters to centimeters, over large $L$ on a scale of up to kilometers,  with very large numbers of coherent quanta.  However, they are still subject to the Planck bound because of new physics described by holographic unification. 
 It is argued below that  universal holographic noise adds another feature not captured by the calculation just presented based on a single quantum in a fixed spacetime:  when the wavefunction is collapsed, the measured displacement  is coherent on scale $\approx L$, even with no physical connection.

\subsection{Holographic Uncertainty in  Black Hole   Evaporation}

Holographic unification was originally motivated by black hole physics, especially  black hole evaporation. The holographic principle\cite{'tHooft:1993gx,Susskind:1994vu,Bousso:2002ju} adopts the view that  the entropy of an  event horizon exemplifies a general property of all null surfaces.
Evaporation offers insights into unification because  the entropy of the black hole event horizon--- an object defined by properties of a pure vacuum solution of classical relativity--- is converted into the entropy of ordinary particle quantum states in a flat spacetime background.    We can use this process to estimate holographic uncertainty of macroscopic position in flat spacetime. Here we sketch the argument without numerical factors of the order of unity, although it should be possible to calibrate the predicted uncertainty with a more careful calculation. 

In Hawking evaporation,  degrees of freedom of a black hole are converted to free particle states. A typical final evaporated state has no black hole; it is a nearly empty vacuum spacetime with about the same number of particles as the number of Planck areas in the original event horizon.
Holographic uncertainty  can be derived directly from the requirement that black hole evaporation obeys quantum mechanical unitarity. The number of degrees of freedom of the position  eigenstates  of particles evaporated from a black hole must agree with the degrees of freedom of the hole.

Consider a black hole of radius $R_S$, in the center of a  giant spherical shell of photographic film of radius $L$.  After a long time, the hole has evaporated. Each evaporated particle has left an image on the film.  In principle the arrival times can be recorded so different arrival times correspond to different states. There are about $N\approx (R_S/\lambda_P)^2$ particles altogether.
At the time a particle evaporates, its wavefunction is close to isotropic (an s-wave state); however,  capturing the images on the film    converts the system to one where the particles are eigenstates of position, or more precisely, emission time and angular direction relative to the hole. Thus the macroscopic system with the recorded images violates the s-wave symmetry of the evaporated wavefunction.  Nevertheless the size of the final Hilbert space must be the same.
The number of states of the evaporated particles cannot be more than the number of states of the hole. This is works out about right  if the distant, flat spacetime has a new  holographic uncertainty in transverse position independently of the properties of the hole.

Think of the hole as covered in Planck size pixels. In thermal evaporation, every time $\approx R_S/c$, the hole is reduced by the typical thermal particle  mass $\approx \hbar/R_Sc$ and shrinks by about one Planck area. The number of different ways to disassemble the hole is the number of ways this can happen $N$ times over. It equals the number of evaporated states if for each particle the number of final directional eigenstates is also $N$. 
%(Every time $R_S/c$ there are $N$ choices of direction. This happens $N$ times before the hole is gone so there are $N^N= \exp{[N\log N]}$ states. The entropy is then $N\log N$. If the difference in ordering is ignored, the entropy is just $\approx N$. The ordering factor cancels in the comparison of the hole with the distant film.)
  Every evaporation time $R_S/c$ there are $N$ choices of direction, and this has to match the number of options for the image location for that particle in the final image.

Suppose that this match is accomplished with  a new uncertainty $\Delta x_L$ in the transverse positions of the particle images, so that the information, or number of degrees of freedom in the film image is $\approx (L/\Delta x_L)^2$.   This refers to  a new relative uncertainty in transverse position over large distances $\approx L$. 
The film can have normally fine-grain pixelation in local measurements; its local physics is not changed. Nearby images, formed at nearly the same time, share the same ``holographic displacement''; nearby displacements are coherent, and there is no local blurring on scale $\Delta x_L$.  However, images formed far apart in time or space have a new indeterminacy or noise in transverse position relative  to each other, by about $\Delta x_L$.

The number of evaporated particle states matches the number of  black hole states if
$(L/\Delta x)^2 \approx (R_S/\lambda_P)^2$. 
This agrees with the black hole entropy as long as
\begin{equation}
\Delta x_L> L  \lambda_P /R_S.
\end{equation}
If $\Delta x_L$ were smaller than this, then the amount of information per evaporated particle would be larger than that available in the hole. It would take more data to specify the final state than is available in the initial state.

Now in addition, the ordinary Heisenberg uncertainty tells us that for the particles of energy  $\hbar c/R_S$, typical for the evaporated particles, there is already a    transverse uncertainty in the image location, $\Delta x > R_S$, that is, the images are locally not sharper than a particle de Broglie wavelength.   Then we get $\Delta x_L^2>\Delta x_L R_S> L\lambda_P$ or
\begin{equation}
\Delta x_L^2> L \lambda_P,
\end{equation}
which is just the holographic uncertainty at distance $L$. It dominates over standard locally-defined Heisenberg uncertainty (that is, it is bigger than the particle wavelength)  for $L> R_S^2/\lambda_P$.

    Indeed this uncertainty does not depend on $R_S$ or any property of the hole, only on $L$, so we conjecture that it is a property of matter position in  flat spacetime.
    Formally we may take $R_S$ to vanish and require that black holes of any size have unitary evaporation.  
    
    Again, the uncertainty is not observable locally--- there is no new ``fuzziness"  in a small patch of the distant film. Rather, the relative angular positions of widely separated images formed  are fuzzy, in a coherent way relative to each other, or relative to the position of the hole. If images form within a time $<< L/c$ of each other  and are physically near each other, they share the same holographic displacement; thus the uncertainty is not observable in any measurement on a scale $<<L$.

    If positions in flat spacetime have this transverse indeterminacy, in addition to the Heisenberg uncertainty, then a black hole of any size  can leave behind a state with the same size Hilbert space it started with.
Since the uncertainty in the distant space does not depend on any property of the hole, it must apply to positions in the spacetime even if there is no hole at all.

 \subsection{Degrees of Freedom  of a Massless Field}

In a certain sense, holographic uncertainty is a general property of any frequency-bounded wave field.

Consider a cubic volume of space with a massless scalar field obeying the linear wave equation with reflecting walls. It is conventional to decompose solutions into discrete sinusoidal modes that fit in the box. Given an upper bound on  wave frequency, the number of modes is proportional to the 3D volume.

When such a field is quantized, the modes are identified as harmonic oscillators and separately quantized. Each one is identified as a degree of freedom.
Thus, the number of degrees of freedom appears to be  proportional to volume. Each mode has a number operator and this is one way to decompose the quantum states of the field.

This view is however misleading.  Field configurations  can also be built from a sum of radiation entering along each face of the cube, with a cutoff in transverse wavenumber corresponding to the maximum frequency.   
The number of degrees of freedom  for propagating waves and particles is thus proportional to surface area, not volume.
The reconciliation of these two conclusions is that the 3D modes are not really independent degrees of freedom. The linear decomposition makes it look that way, and certainly one can build any classical solution by adding linear plane wave mode solutions.  But the 3D plane waves are not independent  quantum degrees of freedom.  A given field configuration can be described by many different linear combinations of plane wave modes.  A more rigorous derivation showing holographic scaling of total entropy in quantum field theory is given in ref.\cite{Srednicki:1993im}.

\section{Wave Picture of Holographic Spacetime}

The macroscopic behavior of holographic unification can be interpreted using a wave model of emergent spacetime.
The relationship between transverse  positions at two times is described by a quantum wavefunction. 

In previously described quantum operator language\cite{Hogan:2007pk}, the transverse positions at different times along a null path are conjugate observables, like normal position and momentum in ordinary quantum mechanics, with  a commutator $[\hat x_1,\hat x_2]= - i ct_{12}\lambda_P$. In a more localized form,
$\label{localcommute}
[\hat x,\hat \theta_{xz}]=-i\lambda_P.
$
The observable operator $\hat x$ represents the center of mass position of mass-energy on a particular wavefront. In a holographic light sheet frame, a state describes the null volume swept out by a sheet, including transverse center of mass position along that null trajectory   in some lab frame.  The variable $\theta_{xz}$ is interpreted as an angular deviation from the classical $z$ axis, the normal defined by null fields in our 3D world.

These Heisenberg-like uncertainty relations give the same result for the wavefunction width as a wave optics calculation.  
The wavefunction  can be computed  in the same way as transverse mutual complex correlation of field amplitudes for radiation of frequency $c/\lambda_P$ in standard wave diffraction theory\cite{Hogan:2008zw}. The width of the mutual complex correlation, the joint wavefunction of transverse position at points separated by $ct$,  increases like $\Delta x\simeq \sqrt{ct \lambda_P}$.

The wave hypothesis  developed here is  based on  the paraxial wave equation, another well studied tool in optics theory.  This   formulation   allows a more direct connection with a macroscopic limit of fundamental theories, such as Matrix theory.  The circumference of a compactified  M dimension is identified with the wavelength of a carrier wave in  the emergent spacetime, and the complex phasor amplitude of deviations from the carrier  is identified as a wavefunction  for spacetime position relative to a classical spacetime. This approach allows us to write down specific eigenstate wavefunctions, and use spatial solutions, such as gaussian beams and wave eigenstates, to compute experimental predictions.

\subsection{Paraxial Wave Equation}

A specific way to formulate the holographic hypothesis is to posit that effective spacetime wavefunctions describing macroscopic position states are solutions  of the paraxial wave equation.  
In an emergent 3D space, a 2D light sheet appears as a wavefront moving at the speed of light. The state is thus naturally described as deviations from the wavefronts of a periodic plane wave. The frequency of the carrier is the fundamental frequency in some given lab frame.

Start with the standard 3D wave equation for a field with a single fixed frequency.
In three dimensions, the 3D wave equation for any field component can be written  as the  modulation of a carrier wave,
\begin{equation}
(\nabla^2 +k^2)E(\vec x)=0.
\end{equation}
Here $E(\vec x)$ is a complex phasor representing the amplitude and phase at each point.
We use Euclidean coordinates $\vec x= x,y,z$ to denote positions in an arbitrary lab rest frame.
 A sinusoidal time dependence is built in, $E\propto \sin(\omega t)$, where $\omega =c k= 2\pi c/\lambda$. In holographic geometry the carrier is at the Planck frequency.

To derive the paraxial wave equation, we express the field in the form 
\begin{equation}
E(x,y,z)=u(x,y,z) e^{-ikz}.
\end{equation}
The field $u$   now describes deviations from a plane wave normal to the $z$ axis. 
In laboratory optics applications, $z$ corresponds the direction of a beam, and $x$ to the width of a beam.  In our holographic application $z$ corresponds to  position in a particular direction  that defines the normal axis of a holographic frame, and $x$ to position in a transverse dimension.
The wave equation for $u$ becomes
\begin{equation}
\partial^2 u/\partial x^2 + \partial^2 u/\partial y^2+\partial^2 u/\partial z^2 - 2 i k \partial u/\partial z = 0.
\end{equation}
The paraxial approximation is to assume that the third term is negligible compared with the others:
\begin{equation}\label{paraxial}
{\partial^2 u\over \partial x^2 } +{ \partial^2 u\over \partial y^2}  - 2 i k {\partial u\over\partial z} = 0.
\end{equation}
This equation is proposed as an effective wave equation governing transverse position states of mass-energy relative to classical spacetime on macroscopic scales, in a small-angle approximation.
Equation (\ref{paraxial}) is the same as the  nonrelativistic Schr\"odinger wave equation, with $z$ replacing time and $-k$ replacing $m/\hbar$.

It should be emphasized that this phenomenological description is not a fundamental theory. The carrier field is not a dynamical physical field, but a calculational tool. It is constructed to represent the holographic behavior in a lab frame; thus, the wavefunction represents the slowly varying parts of the spatial behavior relative to a Planck frequency plane wave. A true carrier field would not be invariant under boosts to another frame, and neither is this; the wavefunctions are frame-dependent. Similarly, the expansion in paraxial coordinates makes sense if the fundamental theory is built on 2D light sheets, even if the actual wavefronts are not the same in a different lab frame. 

This effective theory describes the imperfections in macroscopic geometrical information  compared to the classical limit. It is based on an invariant wave equation, so different observers agree about classical observables. 
 The holographic displacement itself is not invariant: it depends not only on the observer's frame but also on the measurement being made. On the other hand, as noted above the effect is very small and would not have been detected so far in tests of Lorentz covariance.

\subsection{Macroscopic Interpretation of Matrix theory}
 
    The  interpretation of Eq. (\ref{paraxial}) as a wave equation for spacetime also appears to be  natural  in a particular  macroscopic interpretation\cite{Hogan:2008ir}  of Matrix theory\cite{Banks:1996vh}. 
In that theory, the fundamental objects are nine $N\times N$ matrices, $X_i$.    Spatial position operators $\hat x_i\equiv  tr \hat X_i$ refer to the center of mass of a collection of $N$ D0 branes or particles in each of nine spacelike dimensions.  In our interpretation, these are identified as transverse position operators. Seven of the nine spacelike dimensions are  microscopic,  two are macroscopic, and the third macroscopic spacelike dimension emerges holographically.    The  3+1D world has a minimum wavelength determined by the radius $R$ of a compactified $M$ dimension, the only scale in the system;  the circumference $2\pi R$ sets the minimum wavelength in the  emergent spacetime,  the wavelength of the carrier wave in the paraxial wave description. 

The kinematic part of the Banks et al.\cite{Banks:1996vh} Matrix Hamiltonian  can be written
\begin{equation}
\hat H ={R\over 2 \hbar} tr \hat \Pi_i\hat \Pi_i,
\end{equation}
where the $\hat \Pi_i$ denote conjugate operators to $\hat X_i$. In writing this, we ignore all the details of particle dynamics and interactions described by the other terms of the  Matrix theory Hamiltonian.  The holographic uncertainty can be described using  just this kinematic equation.

This leads to a Schr\"odinger  wave equation resembling Eq. (\ref{paraxial}) if we make operator identifications
similar to those in the standard Schr\"odinger wave theory. We can substitute  the light sheet coordinate $z^+\equiv (z+ ct)/2$ for $t$ in the Hamiltonian operator, since for events on a null trajectory, $z^+=ct=z$. For two transverse dimensions $x$ and $y$, we write
\begin{equation}\label{momentum}
tr \hat \Pi_i\hat\Pi_i\rightarrow - \hbar^2(\partial^2/\partial x^2+ \partial^2/\partial y^2),
\end{equation}
\begin{equation}\label{hamilton}
  \hat H\rightarrow i \hbar\partial /c\partial t \qquad {\rm or} \qquad i \hbar\partial /\partial z^+,
\end{equation}
and set $R\rightarrow k^{-1}=\lambda_0/2\pi$. As in ref. \cite{Hogan:2008ir}, we   leave the minus sign in the squared momentum operator, or equivalently, adopt the usual Schr\"odinger imaginary momentum, $ -i\hbar\partial/\partial x$. The   wave equation for M theory wavefunctions in the  transverse dimensions  then becomes:
\begin{equation}\label{Mwave}
{\partial^2 u\over \partial x^2 }   + {\partial^2 u\over \partial y^2 } + {4\pi i \over \lambda_0} {\partial u\over\partial z^+} = 0.
\end{equation}
Apart from a sign change in the last term, this is same as the paraxial wave equation derived above.

\subsection{Quantum Model of Holographic Noise}

In any lab frame, we can   use $\partial/c\partial t$ instead of $\partial/\partial z^+$ in Eq. (\ref{Mwave}) for a quantum description of     position jitter  in each transverse dimension with respect to a particular null direction.  In ref.\cite{Hogan:2008ir} it was suggested that when a null wave is reflected in a transverse direction (say, by a beamsplitter in an interferometer), the phase of the wave is described by a transverse position wavefunction. This wavefunction obeys the 1+1D  Schr\"odinger equation,
\begin{equation}\label{Mwave2}
{\partial^2 u\over \partial x^2 }  + {4\pi i \over c \lambda_0} {\partial u\over\partial t} = 0.
\end{equation}
  Solutions  to Eq. (\ref{Mwave2})  can be expressed as a sum of modes,
\begin{equation} \label{matrixsolution}
u(x,t)=\sum_{k^\perp} A_{k^\perp}\exp -i [ ck^+t \pm k^\perp x  ],
\end{equation}
where the wavenumbers  are related by
 \begin{equation}\label{kperp}
k^\perp= \sqrt{4\pi  k^+/\lambda_0}.
\end{equation}

The macroscopic wavelength  $2\pi/ k^+$ is  determined in practice by the arm length $L$ of an interferometer.  The  observable jitter is only affected by modes at longitudinal  frequencies   above $k^+/2\pi=1/2L$, or $k^+>\pi/L$. The apparatus does not record relative positions on larger scales, so the coefficients in the effective wavefunction of an apparatus vanish for smaller $k^+$.

This wave behavior introduces a new variation or noise about any nominal classical value of $x$. For each mode in the wavefunction, the product $k^\perp x$ varies from 0 to $\pm \pi$.  That means that $x$ itself varies from 0 to $\pm \pi/k^\perp$. Using the dispersion relation (Eq. \ref{kperp})  and the limited range of $k^+$, for a mode with period $2L/c$,  $x$ varies harmonically over a finite interval,
\begin{equation}\label{range}
|x|  < \sqrt{\lambda_0  L/2}.
\end{equation}
Shorter modes are also present, leading to a noisy jitter on all timescales and an effective classical behavior resembling a random walk bounded by the maximum range (Eq. \ref{range}).
Note that $\hbar$ has not been assumed to be unity: it has cancelled out, leaving  the Planckian carrier wavelength $\lambda_0$ as the only scale in the theory. 

   Each wave mode in Eq. (\ref{matrixsolution}) combines two dimensions, one spacelike and one null. 
  Their transverse spacelike variation is much smaller than their macroscopic length, so they resemble families of classical rays in an emergent spacetime.  The null or longitudinal ($k^+$) dimension is  identified with time in a particular frame.  The minimum wavelength $\lambda_0$   in any frame in the 3+1D world derives from the  radius $R=\lambda_0/2\pi$ of the M dimension.   The macroscopic wavelength  $2\pi/k^+$  determines the timescale, and also the spatial coherence scale, for microscopic jitter in the transverse $k^\perp$ direction. The spatial coherence, and the fact that transverse amplitude increases with $L$,  means that transverse motions of order $\approx \sqrt{\lambda_0 L}$ are shared in common over a macroscopic scale of order $\approx L$--- a surprising and distinctive feature that can be exploited in experiments.

In a series of measurements, the position jitters randomly and slowly over an interval of about $\approx \sqrt{\lambda_0 L}$, a bounded random walk.  
In a lab frame, we can think of each mode as a family of parallel rays (or an infinite plane wavefront) with a complex inclination that slowly rocks back and forth with an amplitude $\pm \pi/k^\perp$, with a period $2\pi/k^+$ that also corresponds to an emergent longitudinal wavelength.    A complete solution is a superposition or wavepacket of these modes, and a classical random jitter depends on the mixture of modes. The transverse width of the wavepacket corresponds to the probability distribution of transverse position and the amplitude of the transverse jitter.
It describes a quantum relationship (and uncertainty) for position of the matter in any spacelike slice relative to others.
Random phases in the coefficients $A_{k^\perp}$ lead to  noise in any realized time series, which we are calling holographic noise. 

Measurement of a phase in the lab frame collapses the wavefunction.  The measurement setup--- the orientation of the optical surfaces that interact with with the photon wavefronts--- determines the wavefront orientation for the holographic frame, and consequently which direction is transverse, and which ones are longitudinal. Ideal clocks are synchronized by construction with the longitudinal direction,  the M dimension.     The phase  describing the   $x$  position at a macroscopic separation displays a noise when compared to standard clock wavefronts traveling in the $z^+$ direction.

Normally we think of degrees of freedom as almost all residing in independent modes at the microscopic scale. Interferometers are  exquisitely designed to ignore these and instead measure the envelope wavefunction,  the mean or center of mass position of a vast number of particles, on  a macroscopic scale. 
They exclude from the measured signal as many as possible of  the internal degrees of freedom that could potentially add more noise. The Matrix-theory view of this is that the signal directly monitors the trace of one of the (very large dimensional) fundamental matrices corresponding to the center of mass of a whole body, and indeed of other bodies with no other physical connection. Because these are governed by macroscopically-coherent  modes it is possible  for new types of correlations to arise in the wavefunctions and time behavior of separate systems.

\subsection{Paraxial Representation of  Holographic Uncertainty}

The holographic geometry hypothesis is that  macroscopic wavefunctions of position transverse to a light sheet obey the paraxial wave equation (Eq. \ref{paraxial}, with both transverse dimensions now included), with a fundamental wavelength $\lambda_0$, in terms of the  normal coordinate $z$ in any lab frame:
\begin{equation}\label{paraxialsheet}
{\partial^2 u\over \partial x^2 }  + {\partial^2 u\over \partial y^2 } - {4\pi i \over \lambda_0} {\partial u\over\partial z} = 0.
\end{equation}
The $z$  coordinate represents a position along a null trajectory, the emergent direction in the 3D space. In standard Minkowski space coordinates, it is the same as $t$; however, the wave equation is direction-specific. 

The interpretation of this equation is that it governs the wavefunction for the center of mass position of a system of particles in transverse dimensions $(x,y)$ of a  flat spacetime.  In a particular laboratory frame, $x$ and $z$ are standard Euclidean coordinates. The equation refers to a particular holographic frame with  wavefronts normal to $z$.

This is not an equation of motion of the particles in the standard sense: it refers to the quantum wavefunction of position as limited by the holographic nature of unification.  The holographic quantum behavior described by Eq.(\ref{paraxialsheet}) depends not at all on any properties of the particles, except for their relative position: all the usual physics of the field motion and interaction is in addition to this. In the matrix theory argument leading to the paraxial equation, we   discarded all the terms describing fermionic activity and noncommutative matrix geometries, and  included only the kinematic geometrical terms.  The $x,y$ coordinates are thus interpreted as the center of mass position of all mass and energy at a longitudinal position corresponding to a wavefront.

\subsubsection{Gaussian Beam Solutions as Spacetime Wavefunctions}

A set of useful solutions of the paraxial wave equation from  wave optics can be applied to describe transverse position states on wavefronts at null separations.  They describe beams that fall off transversely with a gaussian profile. These gaussian beams comprise  a one-parameter family of solutions, characterized by a longitudinal distance $z_R$ (called the ``Rayleigh Range'') that physically corresponds to a   radius  of wavefront curvature at the place where the beam  has broadened by a factor of two from its narrowest point. Gaussian beams represent the minimal transverse uncertainty.  For a given set of boundary conditions, they saturate the uncertainty principle, in the sense of having  the smallest transverse variance. We adopt these modes as a working model for estimating the level of irreducible, universal holograhic noise in interferometers.

In a given holographic frame,  the Gaussian  solution to (\ref{paraxialsheet}) can be expressed as\cite{kogelnikli,siegman}
\begin{equation} \label{wavesolution}
u(\rho,z)={w_0\over w (z)}\exp\left[ -iz(2\pi/\lambda_0)- i \phi - \rho^2\left( {1\over w^2 }+ {i \pi\over \lambda_0 R}\right) \right]
\end{equation}
where
\begin{equation}
\rho^2=x^2+y^2
\end{equation}
and
\begin{equation}
\phi= \arctan (\lambda_0 z/\pi w_0^2).
\end{equation}

In optics the quantity $R$ represents the real radius of curvature of the wavefronts.  The real part of the wavefunction displays a Gaussian profile in the transverse directions $x,y$.
The  Gaussian width of the beam varies as 
\begin{equation}
w(z)=w_0\sqrt{1+(z /z_R)^2}= w_0\sqrt{1+\left( {\lambda_0 z\over \pi w_0^2}\right)^2},
\end{equation}
and the   radius of curvature
\begin{equation}
R(z)= z \left[1+\left( {\pi w_0^2\over \lambda_0 z}\right)^2\right]
\end{equation}
The width at the  $z=0$ ``waist'' for a given solution, corresponding to a flat wavefront, is
\begin{equation}\label{waist}
w_0=\sqrt{\lambda_0 z_R/\pi}.
\end{equation}
 Note that $w= \sqrt{2}\sigma_u$, where $\sigma_u$ denotes standard deviation of the  wavefunction $u$.
 .
 
The width of the beam gradually broadens with propagation due to diffraction.  
 Modes  with narrower waists diffract more and broaden more quickly.  
The minimum transverse width at a distance $z$ occurs for the mode with $z=z_R$. Conversely at a distance $z_R$ in either direction from the waist, there is a minimum transverse width, with standard deviation $\sigma_u(z_R)=w_0=  \sqrt{\lambda_0 z_R/\pi}$. 

 A system of spacetime and mass-energy can be put in different states, represented by  solutions characterized by $w_0$ or $z_R$,  by a measurement apparatus.  A small transverse width at the waist implies a larger transverse width at large $z$. For a given $z$ separation between two measurements, there is an optimum state which minimizes the uncertainty in the sum or difference of $x$ for the two measurements.
The Gaussian solution is the least uncertain.

In a theory where states are encoded on light sheets with a characteristic or maximum frequency, the transverse width of the position wavefunction  displays this minimum diffractive uncertainty.  The $x$ observable described by this wavefunction is the position of ``everything''--- the center of mass of all mass-energy on the wavefront.

\section{Statistics of Holographic Noise in  Signals}

Holographic uncertainty (in analogy to Heisenberg uncertainty) can be defined as the minimum of the product of the widths of transverse wavefunctions $u$ at some separation $L$.
The value of the difference of the positions  is indeterminate, leading to  holographic indeterminacy in an instrument, such as an interferometer, that measures such a difference.
Holographic noise comes about because the indeterminacy leads to a random variation in the measured position as a function of time.

\subsection{Setting $\lambda_0$ from Black Hole Physics}

The wave theory is normalized by matching   degrees of freedom    to gravitational physics.
We require the theory to agree with the entropy per transverse lightsheet area from black hole thermodynamics,
\begin{equation}
S_H/A=(4 \lambda_P^2)^{-1},
\end{equation}
where Boltzmann's $k$ is set equal to unity.

This entropy corresponds to one  degree of freedom per $2\lambda_P$ in each transverse direction. 
Standard quantization for a confined scalar particle gives  one degree of freedom per $\lambda$ for each direction. These agree if the effective fundamental wavelength, for macroscopic purposes, is twice the standard definition of the Planck length:
\begin{equation}\label{effective}
\lambda_0\equiv 2\pi/k_0\equiv ct_0 = 2\lambda_P\equiv 2 c t_P.
\end{equation}
Henceforth we will normalize to this value for predictions of observable effects.

This may be an exactly   correct value for all practical purposes, 
since it invokes new physics only via gravitational entropy, which is well calibrated with ordinary entropy using the theory of black hole evaporation. However, this derivation falls short of a rigorous proof, and does not offer a detailed characterization of the actual degrees of freedom.  A better understanding of the fundamental degrees of freedom, and how they relate to gravity, is one goal of an experimental program.

\subsection{Autocorrelation of Displacement}

 We wish to compute the statistical properties of an interferometer signal, a scalar phase as a function of time.  Suppose that the effect of  holographic noise   mimics a classical motion of the beamsplitter as described above.  Denote by $X(t)$ the difference in  position of the beamsplitter along the two arm directions at time $t$ in a lab frame.  
 The time correlation of the classical scalar  $X$ is defined as the limiting average,
 \begin{equation}\label{correlation}
\Xi(\tau)=\lim_{T\rightarrow\infty} (2T)^{-1}\int_{-T}^{T} dt X(t) X(t+\tau)
\end{equation}
In a particular interferometer optical configuration (depending on signal recycling, folded arms, Fabry-Perot arm cavities, etc.),  $\Xi(\tau)$ determines the statistics of the noise in the signal.

 In an interferometer we  interpret $u(x,z)$ as a wavefunction of position and use it to compute the time correlation of the measurements.  We model the wavefunction of the beamsplitter in each direction as the waist of a gaussian beam mode  where the radius $z_R=L$. 
The correlation (Eq. \ref{correlation}) at lag $\tau=0$ is given by the mean square displacement  determined by the  wavefunction (Eq. \ref{wavesolution}) with $z_R=L$, with an extra factor of 2 because we add the variance from two events:
\begin{equation}\label{totalvariance}
\Xi(\tau=0)=2 \int dx x^2 u^*(x, z=z_R=L)u(x, z=z_R=L).
\end{equation}
The standard deviation of each  gaussian $u$ is at the waist is  $\sigma_u=w_0/\sqrt{2}$; for their product, $\sigma=w_0/2$.  The integral is  twice the mean square,
given by $2 w_0^2(z=z_R)/4=w_0^2/2 =z_R\lambda_0/2\pi=L\lambda_0/2\pi$.  
  This effective displacement variance is the same as the quantum limit estimated directly from the standard  uncertainty principle, Eq. (\ref{Deltax}). This calculation normalizes the observable correlation functions of signals.

The effect of the noise mimics a bounded  fluctuation about a nominal classical position. The position distribution has a total variance given by Eq. (\ref{totalvariance}) but is uncorrelated for times separated by more than $2L/c$.  Since the variance increases linearly with scale, we  guess that the time autocorrelation of apparent arm length difference has the approximate form
\begin{equation}\label{smalltau}
\Xi(\tau)=(c  t_0/ 4\pi)[2L-c\tau] , \qquad 0<\tau<2L/c,
\end{equation}
  \begin{equation}\label{largetau}
\Xi(\tau)=   0,  \qquad  \tau> 2L/c.
\end{equation}

These formulas do not capture the correct behavior at very small $c\tau$, less than the beam diameter, say. However, for practical purposes Eqs. (\ref{smalltau}) and (\ref{largetau}) are a prediction of the effective classical behavior.  Time-domain sampling of the signal is predicted to show this correlation.  

The behavior can also be described in  the frequency domain.   The spectrum   $\tilde\Xi(f)$ is given by the Wiener theorem,
%\footnote{I omit the normalization of $(2\pi)^{-1}$, as is conventional in gravitational wave physics.},
\begin{equation}
\tilde\Xi(f)= 2 \int_0^\infty d\tau \Xi(\tau) \cos(\tau\omega),
\end{equation}
where $\omega=2\pi f$.\cite{camp04}
Integration of this formula using Eqs.(\ref{smalltau}) and (\ref{largetau}) gives a prediction for the spectrum of the holographic displacement noise,
\begin{equation}
\tilde\Xi(f)= {c^2t_0\over 2\pi (2\pi f)^2}[1-\cos (f/f_c)], \qquad f_c\equiv c/4\pi L.
\end{equation}

The low frequency limit ($f<<c/2L$) gives a flat spectrum independent of $f$:
\begin{equation}\label{lowfreq}
\tilde\Xi(f)\approx t_0L^2/\pi= 2 t_PL^2/\pi, \qquad f<< c/2L.
\end{equation}
The spectrum at  frequencies above $f_c$ oscillates with a decreasing envelope.  The apparatus size--- the distance over which the sampling occurs--- acts as a high-pass filter; fluctuations from longer longitudinal modes do not enter into the signal.
The exact predicted spectrum may differ at high frequencies, since this description of the apparatus is still based on a  semiclassical   wave model.

\subsubsection{Equivalent Strain  Spectral Density}

A model of an apparatus using the beamsplitter position correlation function (Eqs. \ref{smalltau}, \ref{largetau}) as a description of effective classical motion   allows a prediction of the signal statistics at all frequencies. Current results are generally quoted in terms of equivalent gravitational wave strain, which requires a consideration of the gravitational wave transfer function of an apparatus.

A gravitational wave is a perturbation of the spacetime metric. As emphasized earlier, the holographic effect is not a metric perturbation, but a fluctuation of mass-energy position from a classical geodesic.   A gravitational wave affects interferometer signals by causing a perturbation in the difference of integrated distance traversed by light in the two arms. Holographic noise affects the signals by a change of position transverse to null wavefronts whose phase is being measured. The apparent gravitational wave spectrum of holographic noise depends on the  configuration of the apparatus. In particular, it  changes if the arms of the interferometer are folded, or  include resonant Fabry-Perot cavities that amplify the effect on phase of an arm length perturbation.

In the low frequency limit (Eq. \ref{lowfreq}), the  effective holographic beamsplitter displacement noise  in a folded Michelson interferometer creates the same noise spectrum as an amplitude spectral density of gravitational waves given approximately by:
\begin{equation}\label{apparent}
h(f)\approx  {\cal N}^{-1} \sqrt{\tilde\Xi/L^2} =  {\cal N}^{-1}\sqrt{2t_P/\pi}=  {\cal N}^{-1} 1.84\times 10^{-22}/\sqrt{\rm Hz},
\end{equation}
where ${\cal N}$ is the average number of photon round trips in the interferometer arms.

 The reason for the added factor of ${\cal N}^{-1}$ in Eq.(\ref{apparent})  is that folded arms (as in GEO600) amplify the signal response to a gravitational wave strain, but do not increase the holographic noise in the signal. The holographic displacement of the  beamsplitter and inboard folding mirrors  in physical length units just depend on the actual size of the arms. The folding effectively lengthens the arms for gravitational wave detection by up to a factor of $ {\cal N}$, but it does not change the displacement spectrum or amplify the holographic noise in the signal.  
Thus for a given physical displacement of the beamsplitter, the measured phase displacement corresponds to a gravitational-wave strain   proportional to ${\cal N}^{-1}$ at frequencies below $\approx c/2L{\cal N}$.  

One might worry that the inboard mirrors share the holographic motion of the beamsplitter.  The holographic movements of the inboard mirrors (if they are close to the beamsplitter) are almost the same as those of the beamsplitter. They execute the same bounded random walk about a common classical position. However, those motions are transverse to the beam in each arm so there is no effect on the signal phase from the inboard reflections.  
  
  In GEO600, with ${\cal N}=2$, the estimate in  Eq.(\ref{apparent}) predicts a new noise source, $h=\sqrt{t_P/2\pi}=0.92 \times 10^{-22}/\sqrt{\rm Hz}$, at all measured frequencies.    Indeed if it is real,  holographic noise is currently an important noise source in GEO600 at its most sensitive frequency, around 500 Hz.
 In ref.  \cite{Hogan:2008zw} a similar result was derived, by a calculation based on a wave-optics model similar to that presented here. In that paper however it was erroneously claimed that in the GEO600 power-recycling cavity the predicted slope changes at very low frequencies, below an inverse power-recycling time. In fact the apparent gravitational wave spectrum corresponding to the predicted bounded random walk of the beamsplitter is just flat at low frequencies, as in Eq. (\ref{apparent}). In addition, the numerical factor in ref.  \cite{Hogan:2008zw} was different, $h=\sqrt{t_P/2}$ instead of $h=\sqrt{t_P/2\pi}$, so the predicted noise here is slightly less. The  current calculation takes into account the detailed profile of the gaussian mode solution, Eq. (\ref{waist}),  which is likely to be a more physically realistic model of instrument/spacetime wavefunction,   and should be taken as a more reliable calculation than the earlier one of the actual minimum level of noise.  
    
GEO600 is more sensitive than LIGO to holographic noise, even though it is less sensitive to gravitational waves. 
 The equivalent strain noise  predicted in LIGO is well below current limits due to its  Fabry-Perot arm cavities, which have  a high finesse corresponding to  ${\cal N}\approx 10^2$.  (Note that by contrast,  LIGO data already rule out  other models  predicting Planck-amplitude noise in the metric \cite{AmelinoCamelia:1999gg,AmelinoCamelia:2001dy,Ng:2004xr}).     
 
The equivalent strain formula (Eq. \ref{apparent}) is only valid at low frequencies.   
For a folded system, the   amplification of the effect on the signal  for gravitational waves decreases above a frequency $\approx c/2L{\cal N}$, since there are fewer roundtrips per wave cycle.  Thus the apparent gravitational wave spectrum actually rises from there up to a frequency $\approx c/2L$, above which it is about the same as an unfolded system. At $f\approx c/2L$, both types of system have an effective spectral amplitude $h\approx \sqrt{t_P}$, which just corresponds to  $\Delta L\approx \sqrt{Ll_P}$.     At frequencies above $\approx c/2L$, the apparent strain noise spectrum  turns over to $h(f)\propto  (c /f L) \sqrt{t_P}$.
For example, holographic noise
should appear in LIGO as an apparent (but illusory) stochastic gravitational-wave background at high frequency, with a  maximum amplitude near the free spectral range of the system. 

\subsection{Cross Correlation  of Two Interferometers}

An experiment designed to provide convincing evidence for or against this interpretation of holographic unification could include more than one interferometer. Two separate interferometers, with no physical connection aside from inhabiting the same holographic spacetime volume, should nevertheless show correlated holographic noise. This feature can be used to design an experiment with clear holographic signatures in the signal.

Coherent jitter is perhaps the most specific and radical aspect of holographic uncertainty.  Since the effective motion is coherent in the transverse direction, there is little effect on local physical observables. Without coherence, there would be locally-observable jitter dependent on arbitrarily distant observers. Since the jitter is coherent, it is only observable by measurements that compare position nonlocally, over large spacetime volumes.  At the same time, aligned measurements covering nearly the same spacetime volume show nearly the same jitter.

The detailed correlation of two interferometers is suggested by the matrix theory interpretation and wave solution (Eq. \ref{matrixsolution}). The noise arises from a superposition of modes with random phases.  The particular  pattern of a signal is determined by the phases of the modes, which do not depend on transverse spatial  location. For each mode the phase in the transverse displacement (on a scale $\lambda^\perp \approx 2\pi/k^\perp$ in Eq. \ref{kperp}) is coherent over the macroscopic longitudinal/temporal wavelength ($\lambda^+\approx 2\pi/k^+$).  On larger scales, the coherence disappears.

 The details of a particular observed time series only reveal themselves in the course of the decoherence or collapse of the wavefunction, that is, measurements.  Since phases do not depend on spatial position, measurements are consistent (in the paraxial approximation) at all points in a given wavefront, at a given time.
The correlation is expected insofar as the optical elements of the two interferometers are described by different  sets of D0 branes on the same light sheets at the same time. They all share the same holographic wavefunction for their collective transverse position, the trace of the matrix including all the branes.  The position variable represents the center of mass position without reference to other physical connections.  

The effect of a  measurement on collapsing a wavefunction is confined to its future light cone.  Also, a phase measurement can only depend on events in its past light cone.  Thus for measurements to be correlated, two relevant light cones must overlap: the future light cone of one reflection, and the past light cone of the second reflection with which it is compared.  Hence   to obtain correlated signals the light paths (as in Figure \ref{lightcone}) must enclose a common spacetime volume.

The cross correlation has the effect of reducing the variance of the difference of signals in two interferometers, $A$ and $B$. Writing $X_A$ and $X_B$ for the effective arm length difference in each, 
\begin{equation}
\langle (X_A-  X_B)^2\rangle= 2( \langle X_{A,B}^2\rangle- \langle X_A X_B\rangle)
=2\Delta X^2 - 2\langle X_A X_B\rangle
\end{equation}
For nearly collocated interferometers and zero time lag, the left side nearly vanishes.  The cross correlation $\langle X_A X_B\rangle$ now appears not as an excess noise, but  as a measurable signal that can be distinguished  from other independent sources of noise, for which $\langle X_A X_B\rangle=0$.

In general, the cross correlation is a function of time lag $\tau$.
The cross correlation of the scalars  $X_A,X_B$ is defined as the limiting average,
 \begin{equation}\label{xcorrelation}
\Xi(\tau)_\times=\lim_{T\rightarrow\infty} (2T)^{-1}\int_{-T}^{T} dt X_A(t) X_B(t+\tau).
\end{equation}

The cross correlation signal can be modulated: it gradually disappears as the interferometers are separated in space. For close to zero displacement, cross correlation power is added linearly  when aligned laser beam wavefronts are traveling in both arms of both interferometers at nearly the same time.  If the  in-plane displacement  exceeds the arm length in either direction, the correlation disappears. This is the consequence of causality: correlations only arise if causal spacetime volumes defined by light paths in the two machines overlap. Using these criteria, we can make a guess at the approximate cross correlation behavior for displacements along the arm directions.

Consider two aligned interferometers, both  with arms of length $L$  parallel to the $x$ and $y$ axes. Suppose one is placed vertically above the other, with  vertical displacement $\Delta L_z$.  If  $\Delta L_z<<L$, the holographic position displacements of the two beamsplitters  are almost the same, differing from each other by only $\approx \sqrt{\Delta L_z \lambda_P}$. 

Consider the holographic frame with wavefronts aligned normal to the $z$ axis, that is, parallel to the $xy$ plane containing the two arms. The positions of the two instruments at null separation (at times differing by $\Delta L_z/c$ in the lab frame)  are described by data on the same holographic wavefront.  The uncorrelated $x,y$ holographic displacement between the two at any time along each axis is then given by the transverse uncertainty in this frame, which is only $\Delta x\approx \Delta y\approx \sqrt{\Delta L_z \lambda_P }$; the rest of the apparent motion must be correlated. Therefore, for   $\Delta L_z<<L$, the cross  correlation must be nearly perfect;  the spacetime wavefunctions for the differences of  $x$ and $y$ between the two systems are very narrow compared to the total wavefunction width corresponding to $\approx \sqrt{ L \lambda_P}$. This follows as long as the two transverse degrees of freedom are independent in any holographic frame.  
The coherence is connected to the paraxial approximation used to estimate the indeterminacy.
If $\Delta L_z$ is larger than $L$,  the coherence must vanish for the emergent 3D system to obey causality in all directions (that is, in any frame).

For  small displacements of two aligned interferometers offset along either arm by $\Delta L$, these considerations suggest a cross correlation of the form
\begin{eqnarray}\label{small}
\Xi_{\times}(\tau)&\approx  & (ct_0/ 4 \pi) (2L-2\Delta L- c\tau),    \qquad   0<c\tau<2L-2\Delta L\\
&=  & 0,  \qquad  c\tau>2L-2\Delta L.
\end{eqnarray}
These formulas likely describe the decorrelation accurately for small displacements,  but the framework here does not suffice to treat large displacments, or to compute correlations for  general displacements in three dimensions.  Similarly, the correlation between two interferometers decreases like $\cos(\theta)$ if they are misaligned with each other by angle $\theta$.  It vanishes if they are misaligned by more than 90 degrees, since they probe independent spacetime volumes. 

Indeed in general, correlations must vanish between regions of spacetime that are causally disconnected. This implies an upper limit to the cross correlation of two interferometers. Define $L'$ as the size of the largest fully enclosed arm pair:  that is, an arc of this size drawn centered on  one beamsplitter between its two arms is fully enclosed in an arc of size $L$ centered on the other and between its arms.
If maximal correlations arise from the in-common causal spacetime volume defined by reflection pairs in the two machines, we obtain
\begin{equation}
\Xi_{\times}(\tau)< (ct_0/ 4 \pi) (2L').
\end{equation}
This is consistent with the estimate for small displacements (Eq. \ref{small}).
The frequency spectrum of the  cross correlation signal is given as above using the Wiener formula.  For $\Delta L> 0$,  the  noise power is reduced at all frequencies. The low frequency limit  becomes
\begin{equation}
\tilde\Xi_{\times}(f)\approx   t_0 L^2  [1-(\Delta L/L)]/ \pi, \qquad f<< c/2L.
\end{equation}
%\begin{equation}
%\tilde\Xi(f)\approx (At_0c^2/ f^2)  [1-(L_{\off}/L)], \qquad f>> c/2L.
%\end{equation}

\section{Experimental Program }

Gravitational wave interferometers, while extraordinarily sensitive, are not designed to obtain a clean signature of holographic noise. Either a detection or an upper limit requires modeling and precise calibration of other noise sources.   

A new experiment is therefore motivated to test the holographic hypothesis. Such an experiment  could exploit the  predicted signal correlation between two nearly collocated interferometers. For two aligned interferometers with a fractionally small displacement, a cross correlation in the holographic noise displacement is robustly predicted on general theoretical grounds.  The two interferometers  can be in separate cavities and Faraday isolated to exclude other sources of in-common continuum noise in the relevant band. The dominant broadband noise source, photon shot noise, is uncorrelated between the two and averages in time to zero, while the holographic noise is in common and averages to a definite known predicted value. This allows an experiment on a much smaller scale (and higher frequencies) than interferometric gravitational wave detectors, even though the photon shot noise dominates by a large factor at high frequencies.  Most significantly, it allows a design where the holographic noise yields a positive and distinctive signal.  The correlated signal can be modulated to decrease in a predictable way as the interferometers are separated or misaligned.  

Consider a pair of adjacent and aligned power-recycled interferometers.  The integration time required for the cross correlation from  holographic noise to equal the photon shot noise for a sampling interval  $\tau_{sample}=2L/c$ is\cite{weissconcept}
 \begin{equation}
t_{\gamma\times}\approx \tau_{sample} [\Delta \phi_n^2/\Delta\phi_H^2]^2,
\end{equation}
where the Poisson phase noise is
\begin{equation}
\Delta \phi_n^2\equiv (\dot N\tau_{sample})^{-1}= h_P c^2/2 P_{opt}L\lambda_{opt}.
\end{equation}
Here $N$ is the number of photons, $P_{opt}$ is the cavity power at the beamsplitter, $\lambda_{opt}$ the   wavelength, $h_P$ is Planck's constant, and $L$ is the arm length. Using the estimate above (Eq. \ref{Deltaphase}) for $\Delta\phi_H$   this becomes
\begin{equation}
t_{\gamma\times}\approx  500\  {\rm sec}\  (P_{opt}/ 2000 {\rm w})^{-2}(L/40{\rm m})^{-3}.
\end{equation}
 It appears that an experiment on a modest scale  can  test the holographic noise signatures predicted here.

An experiment that reaches this level of position sensitivity should provide informative results about physics at the Planck scale. A null result excluding cross correlated noise in two systems, in contradiction to these predictions, will imply  that the Planck information flux  bound does not apply to macroscopic position states.  A positive result of course opens up a whole field of followup experiments that will illuminate  the way that Planck-scale quantum physics maps onto the macroscopic, quasi-classical world.

\acknowledgements
I am grateful for discussions and correspondence with many colleagues, including particularly A. Chou,  H. Grote, G. Gutierrez,  S. Hild, M. Jackson, C. Laemmerzahl, H. L\"uck, J. Lykken, G. Mueller, Y. Ng, M. Perry, B. Schutz, J. Steffen, S. Tarabrin, S. Waldman,  R. Weiss, S. Whitcomb, and G. Woan,  and for the hospitality of the Aspen Center for Physics. This work was supported by the Department of Energy, and by NASA grants NNX08AH33G  at the University of Washington and  NNX09AR38G at the University of Chicago.


\begin{thebibliography}{}

 
 \bibitem{wigner}
E. P. Wigner, ``Relativistic Invariance and Quantum Phenomena'', Rev. Mod. Phys. {\bf 29}, 255 (1957)

\bibitem{salecker}
H. Salecker \& E. P. Wigner, ``Quantum Limitations of the Measurement of Space-Time Distances'', Phys. Rev. {\bf 109}, 571 (1958)

\bibitem{kempf}
A. Kempf,  ``Information-Theoretic Natural Ultraviolet Cutoff for Spacetime'', Phys. Rev. Lett. {\bf 103},  231301 (2009)

\bibitem{mattingly}
D. Mattingly, 
%"Modern Tests of Lorentz Invariance", 
Living Rev. Relativity 8,  (2005),  5. 


%\cite{'tHooft:1993gx}
\bibitem{'tHooft:1993gx}
  G.~'t Hooft,
  ``Dimensional reduction in quantum gravity,''
  in ``Conference on Particle and Condensed Matter Physics (Salamfest)'',
  edited by A. Ali, J. Ellis, and S. Randjbar-Daemi (World Scientific, Singapore, 1993),
  arXiv:gr-qc/9310026.
  %%CITATION = GR-QC/9310026;%%

%\cite{Susskind:1994vu}
\bibitem{Susskind:1994vu}
  L.~Susskind,
  ``The World As A Hologram,''
  J.\ Math.\ Phys.\  {\bf 36}, 6377 (1995)
% [arXiv:hep-th/9409089].
  %%CITATION = JMAPA,36,6377;%%
  
   %\cite{Bousso:2002ju}
\bibitem{Bousso:2002ju}
 R.~Bousso,
 ``The holographic principle,''
 Rev.\ Mod.\ Phys.\  {\bf 74}, 825 (2002)
% [arXiv:hep-th/0203101].
 %%CITATION = RMPHA,74,825;%%
 

%\cite{Banks:1996vh}
\bibitem{Banks:1996vh}
  T.~Banks, W.~Fischler, S.~H.~Shenker and L.~Susskind,
  ``M theory as a matrix model: A conjecture,''
  Phys.\ Rev.\  D {\bf 55}, 5112 (1997)
 % [arXiv:hep-th/9610043].
  %%CITATION = PHRVA,D55,5112;%%

  
  %\cite{Hogan:2007pk}
\bibitem{Hogan:2007pk}
  C.~J.~Hogan,
  ``Measurement of Quantum Fluctuations in Geometry''
  Phys. Rev. D 77, 104031 (2008),
  %arXiv:0712.3419 %[gr-qc].
  %%CITATION = ARXIV:0712.3419;%%
  
  %\cite{Hogan:2008zw}
\bibitem{Hogan:2008zw}
  C.~J.~Hogan,
  ``Indeterminacy of Quantum Geometry''
  Phys Rev D.78.087501 (2008),
  % arXiv:0806.0665% [gr-qc].
  %%CITATION = ARXIV:0806.0665;%%

%\cite{Hogan:2008ir}
\bibitem{Hogan:2008ir}
  C.~J.~Hogan and M.~G.~Jackson,
  ``Holographic Geometry and Noise in Matrix Theory,'' Phys. Rev. D.79.124009 (2009)
  % http://link.aps.org/abstract/PRD/v79/e124009
%arXiv:0812.1285 [hep-th].
  %%CITATION = ARXIV:0812.1285;%%

  
  

  
    %\cite{Ellis:1983jz}
\bibitem{Ellis:1983jz}
  J.~R.~Ellis, J.~S.~Hagelin, D.~V.~Nanopoulos and M.~Srednicki,
  ``Search For Violations Of Quantum Mechanics,''
  Nucl.\ Phys.\  B {\bf 241}, 381 (1984).
  %%CITATION = NUPHA,B241,381;%%
  
  %\cite{AmelinoCamelia:1999gg}
\bibitem{AmelinoCamelia:1999gg}
  G.~Amelino-Camelia,
  ``Gravity-wave interferometers as probes of a low-energy effective  quantum
  gravity,''
  Phys.\ Rev.\  D {\bf 62}, 024015 (2000)
  %[arXiv:gr-qc/9903080].
  %%CITATION = PHRVA,D62,024015;%%
  
   %\cite{AmelinoCamelia:2001dy}
\bibitem{AmelinoCamelia:2001dy}
  G.~Amelino-Camelia,
  ``A phenomenological description of quantum-gravity-induced space-time
  noise,''
  Nature {\bf 410}, 1065 (2001)
  %[arXiv:gr-qc/0104086].
  %%CITATION = NATUA,410,1065;%%

     
%\cite{Schiller:2004tf}
\bibitem{Schiller:2004tf}
  S.~Schiller, C.~Laemmerzahl, H.~Mueller, C.~Braxmaier, S.~Herrmann and A.~Peters,
   ``Experimental limits for low-frequency space-time fluctuations from
  ultrastable optical resonators,''
  Phys.\ Rev.\  D {\bf 69}, 027504 (2004)
 % [arXiv:gr-qc/0401103].
  %%CITATION = PHRVA,D69,027504;%%

  %\cite{Ng:2004xr}
\bibitem{Ng:2004xr}
  Y.~J.~Ng,
  ``Quantum foam and quantum gravity phenomenology,''
  Lect.\ Notes Phys.\  {\bf 669}, 321 (2005)
  %[arXiv:gr-qc/0405078].
  %%CITATION = LNPHA,669,321;%%
  
   %\cite{Smolin:2006pa}
\bibitem{Smolin:2006pa}
  L.~Smolin,
 ``Generic predictions of quantum theories of gravity,''
  arXiv:hep-th/0605052.
  %%CITATION = HEP-TH/0605052;%% 
  
  \bibitem{fermi2009}
  A. A. Abdo et al., Nature {\bf 462}, 331-334,  doi:10.1038/nature08574 (2009)
  
  
  \bibitem{zurek}
  W. H. Zurek, ``Decoherence,  einselection, and the quantum origin of the classical'', Rev. Mod. Phys. {\bf  75}, 715 (2003)
  
  \bibitem{braginsky}
V. B. Braginsky and F. Ya. Khalili, ``Quantum Measurement'',  Cambridge:  University Press (1992)

  

  
  \bibitem{caves81}
  C. ~Caves, 
  ``Quantum-Mechanical Noise in an Interferometer'', 
  Phys. Rev. D {\bf 23}, 1693 (1981)
  
%\cite{Srednicki:1993im}
\bibitem{Srednicki:1993im}
  M.~Srednicki,
  ``Entropy and area,''
  Phys.\ Rev.\ Lett.\  {\bf 71}, 666 (1993)
 % [arXiv:hep-th/9303048].
  %%CITATION = PRLTA,71,666;%%

  
  \bibitem{kogelnikli}
  H. Kogelnik, T. Li, ``Laser Beams and Resonators'', Appl. Opt. 5, 1550 (1966)
  

  \bibitem{siegman}
A. E.  Siegman, {\it Lasers},  Sausalito: University Science Books (1986).

\bibitem{camp04}
J. B. Camp, N.J. Cornish, ``Gravitational Wave Astronomy'', Ann. Rev. Nuc. Part. Sci. 54,  525 (2004)

\bibitem{weissconcept}

R. Weiss, ``Concept for an interferometric test of Hogan's quantum geometry hypothesis'', 2/10/2009.
  
    
  
\end{thebibliography}
\end{document}